# CONTROL OF THE MEAN NUMBER OF FALSE DISCOVERIES, BONFERRONI AND STABILITY OF MULTIPLE TESTING

BY ALEXANDER GORDON, GALINA GLAZKO, XING QIU
AND ANDREI YAKOVLEV[1]

*University of Rochester and University of North Carolina at Charlotte, University of Rochester, University of Rochester and University of Rochester*

The Bonferroni multiple testing procedure is commonly perceived as being overly conservative in large-scale simultaneous testing situations such as those that arise in microarray data analysis. The objective of the present study is to show that this popular belief is due to overly stringent requirements that are typically imposed on the procedure rather than to its conservative nature. To get over its notorious conservatism, we advocate using the Bonferroni selection rule as a procedure that controls the per family error rate (PFER). The present paper reports the first study of stability properties of the Bonferroni and Benjamini–Hochberg procedures. The Bonferroni procedure shows a superior stability in terms of the variance of both the number of true discoveries and the total number of discoveries, a property that is especially important in the presence of correlations between individual $p$-values. Its stability and the ability to provide strong control of the PFER make the Bonferroni procedure an attractive choice in microarray studies.

**1. Introduction.** A recent explosion of statistical publications dealing with multiple significance tests has been triggered by the needs of new high throughput technologies in biology such as gene expression microarrays [Dudoit, Shaffer and Boldrick (2003)]. This voluminous literature has been focused on various alternatives to the family-wise error rate (FWER) controlling procedures such as the classical Bonferroni method, the latter having been considered as too conservative for practical purposes.

Received November 2006; revised January 2007.
[1]Supported in part by NIH/NIGMS Grant GM075299 and Johnson & Johnson Discovery Concept Award.
Supplementary material available at http://imstat.org/aoas/supplements

The Bonferroni method was improved by Holm (1979) who proposed a step-down multiple testing procedure (MTP) that has more power but still controls the FWER at the same level. Furthermore, the Holm procedure is known to have strong optimality properties [Lehmann and Romano (2005a), Chapter 9]. Another attempt to gain more power by utilizing the dependence between test-statistics is due to Westfall and Young (1993) who designed a step-down resampling algorithm that provides strong control of the FWER and is consistent (i.e., the FWER approaches its nominal level in large samples) under the condition of subset pivotality. However, the resultant gain from both improvements is quite small, especially when controlling the FWER at a low level. Whenever the FWER is small and the number of hypotheses to be tested is large (e.g., of order $10^4$ as in microarray studies), the ability of any FWER controlling MTP to detect false null hypotheses is inevitably limited. This observation drove statisticians to explore various possibilities with less stringent criteria for guarding against Type 1 errors.

The quest for less conservative MTPs has resulted in many new concepts of error rate, such as the false discovery rate (FDR) [Benjamini and Hochberg (1995), Yekutieli and Benjamini (1999), Benjamini and Hochberg (2000), Reiner, Yekutieli and Benjamini (2003)] and its local version [Efron (2003)], positive FDR [Storey (2003)], generalized FWER [Victor (1982), Dudoit, van der Laan and Pollard (2004), Korn et al. (2004)], tail probabilities for the proportion of false positives [Dudoit, van der Laan and Pollard (2004), Lehmann and Romano (2005b)] and some others. In particular, the beautiful mathematical idea behind the Benjamini–Hochberg (BH) procedure, proposed as a method for controlling the FDR, has attracted considerable attention of statisticians working in the field of multiple testing. Several attempts have been made [Benjamini and Hochberg (2000), Reiner, Yekutieli and Benjamini (2003), Storey, Taylor and Siegmund (2004)] to further increase the "overall average power" within the concept of the FDR. The empirical Bayes method [Efron (2003, 2004), Efron et al. (2001)], which is based on another elegant mathematical idea, serves essentially the same purpose. The two approaches are in a certain sense closely related as discussed in Efron (2003). The motivation for introducing these new concepts and associated MTPs has been, at least in part, the necessity to overcome the excessive conservatism of the FWER controlling MTPs, including the Bonferroni procedure, in the presence of a large number of hypotheses.

Numerous methodological publications on this subject have successfully reached out to practitioners. The original BH step-up procedure and its more liberal versions are frequently confronted with the Bonferroni method and the latter is invariably declared a loser in such comparisons. As a result, the Bonferroni procedure has been effectively disqualified and its practical application has been largely discouraged. It is now difficult to find a published microarray study containing no claim that the Bonferroni method of guarding against Type 1 errors is overly conservative, thereby justifying the need for an FDR controlling procedure in analysis of a specific data set.

It is the intent of the present paper to show that the notorious conservatism of the Bonferroni procedure is a misconception stemming from the traditionally conservative choice of its parameter rather than from any solid evidence of its conservative nature per se. We report the results of a comparative study of the Bonferroni and the BH procedures to show that the outcomes of both procedures are highly correlated when the requirements imposed on their error rates become comparable. To reveal this fact, the threshold parameters of the two procedures need to be properly adjusted so that the comparison of their performance becomes fair. This in turn calls for an extension of the Bonferroni method by focusing on the mean number of false discoveries rather than on the probability of at least one false discovery as a pertinent measure of the Type 1 error rate. The Bonferroni procedure thus interpreted is compared with its natural rival (BH procedure) in terms of random outcomes of multiple tests. The present paper reports the first study of the variability of random outcomes of testing in conjunction with the Bonferroni and BH procedures and this is its main thrust.

**2. An extended interpretation of the Bonferroni method.** The Bonferroni procedure with parameter $\alpha$ ($0 < \alpha < 1$), henceforth denoted by $Bonf^\alpha$, rejects all hypotheses $H_i$, $i = 1, \ldots, m$, whose $p$-values satisfy the inequality $p_i \leq \alpha/m$. The procedure controls the FWER, defined as the probability of one or more false rejections, at level $\alpha$, thereby guaranteeing the probability of rejecting at least one true hypothesis to be less than or equal to $\alpha$ for an arbitrary joint distribution of $p$-values. Another measure of the abundance of Type 1 errors is the per family error rate (PFER), defined as the expected number of false rejections. As noted by Tukey [Tukey (1953)], who introduced the concepts of FWER and PFER, the two error rates are almost indistinguishable when both of them are small, while FWER $\leq$ PFER in the general case. $Bonf^\alpha$ controls the PFER at level $\alpha$ and, consequently, it controls the FWER at the same level. That $Bonf^\alpha$ controls the mean number of Type 1 errors is a simple and well-known fact [Lee (2004)] and yet this procedure has always been perceived only as an FWER-controlling procedure in practical applications.

REMARK 1. Note that the PFER is related to the per comparison error rate (PCER) via the formula PFER $= m$PCER.

We suggest that the current view of the Bonferroni procedure be changed by switching the main focus to its ability to control the PFER. Adopting this extended interpretation would eliminate the requirement that $\alpha$ be much smaller than 1. The latter requirement is essential if $\alpha$ is interpreted as an upper bound for the probability of a rare event. By contrast, if $\alpha$ is interpreted as an upper bound for the *expected number* of false rejections, this requirement becomes irrelevant and the parameter $\alpha$ may be even greater than 1. To make a distinction between the two interpretations, we introduce the notation $Bonf^\gamma$ for the Bonferroni procedure that controls the PFER at the nominal level $\gamma$, which can be any number between 0 and $m$. Since this is essentially the old Bonferroni procedure, we see no reason to change its name when allowing for a wider range of its parameter values. Therefore, we will call $Bonf^\gamma$ the extended Bonferroni procedure. The MTP $Bonf^\gamma$ controls the PFER at level $\gamma$ under any dependence between $p$-values. More precisely, $\text{PFER} \leq (m_0/m)\gamma$, where $m_0$ is the number of true null hypotheses among the $m$ hypotheses to be tested. Indeed, let $\mathcal{T}$ be the set of indices of true hypotheses and $\zeta$ the number of false rejections, then the expectation of $\zeta$ is

$$\mathbb{E}\zeta = \mathbb{E}\sum_{i\in\mathcal{T}} I_{\{P_i \leq \gamma/m\}} = \sum_{i\in\mathcal{T}} \Pr\{P_i \leq \gamma/m\} \leq m_0 \frac{\gamma}{m} \leq \gamma,$$

where $P_i$, $i=1,\ldots,m$, are observed $p$-values and $I_A$ denotes the indicator function of the event $A$. It is clear that $\text{PFER} = (m_0/m)\gamma$ if all $p$-values associated with true null hypotheses are uniformly distributed on $[0,1]$, which is a regular case for continuous data. It follows that $Bonf^\gamma$ controls the FWER at level $[(m_0/m)\gamma] \wedge 1$.

REMARK 2. The procedure $Bonf^\gamma$ with $\gamma$ being not necessarily small has already been discussed in the literature. A single-step procedure considered by Lehmann and Romano (2005b) is equivalent to $Bonf^{k\alpha}$, but it is interpreted as a procedure controlling a generalized familywise error rate ($k$-FWER), defined as the probability of $\geq k$ false rejections, at level $\alpha$. Hence, the focus is still on (small) probabilities rather than expectations. Korn et al. (2004) also mentioned a procedure which is essentially identical to $Bonf^\gamma$ with $\gamma = 10$, but their main focus was on concepts of error rate other than the PFER.

By way of illustration, consider the procedure $Bonf^{0.6}$. When interpreted in the traditional way, this procedure guarantees that, with probability 0.4, there are no false positives. This information is of little utility because the number 0.4 is not close to 1. On the other hand, the requirement that the expected number of false positives should not exceed 0.6 (in the presence of 40000 hypotheses, say) is still quite stringent, whereas a gain in power (relative to $Bonf^{0.05}$, say) can be substantial [Korn et al. (2004)].

**3. Study design and performance indicators.** We generated two sets of simulated data to study the performance of different MTPs in a situation where the "true" and "false" null hypotheses were known. In these simulations each set consists of 500 independently generated pairs of samples of equal sizes. Each sample includes $n = 43$ realizations of a random vector $\mathbf{X}$ with log-normal marginal distributions. The components of $\mathbf{X}$ represent expression levels of 1255 genes, while each realization of $\mathbf{X}$ represents a single array. To model the presence of differentially expressed genes in two-sample comparisons, the mean log-expression of the first 125 genes is set to be equal to 1 in one sample and to 0 in another. The variance of log-expressions is kept equal to 1 in both samples. The log-expressions of the remaining 1130 genes in both samples are generated from a standard normal distribution. The first set of simulated data, denoted by SIM43, represents an ideal case where expression levels of all the genes are stochastically independent. The second set, denoted by SIM43CORR, is generated from a joint log-normal distribution with exchangeable correlation structure as described in [Qiu, Klebanov and Yakovlev (2005)]. In this set of correlated data, all pairwise correlation coefficients are set equal to $\rho = 0.4$, which is deemed quite moderate in view of its typical values estimated from actual microarray data [Almudevar et al. (2006)]. The marginal distributions of log-expressions are identical to those for SIM43. Since our focus is only on proof of principle, the choice of $n$ and $\rho$ is of little consequence to the objectives of this study. We also conducted simulations with an increased effect size with log-expression levels of 125 "different" genes generated from a normal distribution with mean 2 and unit variance. The results are similar and we do not present them here.

We used the standard $t$-test throughout the study. The two MTPs under comparison are the Bonferroni $Bonf^\gamma$ and the Benjamini–Hochberg procedure with parameter $\beta$. The latter procedure, denoted by $BH^\beta$, is known to control the FDR defined as the expected proportion of false discoveries among all discoveries. More specifically, let $R$ be the total number of rejected hypotheses and $V$ the number of true nulls among them. Then the FDR is the expectation of the random variable $\eta$: $\eta = V/R$ if $R > 0$, $\eta = 0$ if $R = 0$. To make the two procedures comparable, we adjust the levels $\gamma$ and $\beta$ so that the nonparametrically estimated true (actual) values of either PFER or FDR, denoted by $\widehat{PFER}$ and $\widehat{FDR}$, respectively, become approximately equal for both procedures. This is done in our simulations by finding such values of $\gamma$ and $\beta$ for which the levels of either $\widehat{PFER}$ or $\widehat{FDR}$ are roughly the same when both procedures are repeatedly applied to 500 simulated samples with their results being averaged to produce the mean values of the said estimators. To generate correspondence tables for $\widehat{PFER}$ and $\widehat{FDR}$ values, we first form a grid of values for the parameter $\gamma$. This grid is not uniform: it has a finer partition at the low end, while tending to be coarser for larger possible values of $\gamma$. The thresholds were chosen as follows: $\gamma_i = 0.01, 0.02, \ldots, 1$ for $i = 1, 2, \ldots, 100$, $\gamma_i = 1.1, 1.2, \ldots, 10$ for $i = 101, 102, \ldots, 190$ and $\gamma_i = 11, 12, \ldots, 100$ for $i = 190, 191, \ldots, 280$. In an effort to provide a more uniform accuracy of the correspondence table, we formed a pertinent nonlinear grid of $\beta$'s as follows: $\beta_j = \frac{\gamma_j}{a+\gamma_j}$ for $j = 1, 2, \ldots, 280$, where $a = 125$ is the number of true alternatives. Each MTP is run at each threshold level to obtain the observed PFER and FDR from each of the 500 samples, yielding the estimates $\widehat{PFER}$ and $\widehat{FDR}$ as the corresponding sample means over the simulations. Then a pair of entries in the correspondence table can be found to indicate those $\gamma_i$ and $\beta_j$ on the two grids that provide approximately the same level of either $\widehat{PFER}$ or $\widehat{FDR}$. For example, when starting with $BH^{\beta_j}$ at different values of $\beta_j$ and using $\widehat{FDR}$ as the equalizer, we define parameters $\gamma_j^*$ for the corresponding Bonferroni procedure as

$$\gamma_j^* = \operatorname*{arg\,min}_{1 \le i \le 280} |\widehat{FDR}_{BH^{\beta_j}} - \widehat{FDR}_{Bonf^{\gamma_i}}|, \qquad j = 1, \ldots, 280.$$

A similar algorithm is designed to equalize the estimated PFER, in which case it is $Bonf^{\gamma_i}$ that produces a set of the estimates $\widehat{PFER}_{Bonf^{\gamma_i}}$ for finding the adjusted thresholds $\beta_i^*$ for $BH^{\beta_i}$.

The adjusted parameters $\gamma_j^*$ and $\beta_i^*$ were used to run both competing MTPs on an independent control set of 500 samples generated in exactly the same way as described above. One performance indicator for the procedures thus adjusted was the standard deviation of the number of true discoveries—this is the most important characteristic related to the power of a given MTP, providing the corresponding mean value is fixed. The overall stability of each of the two MTPs, characterized by the standard deviation of the total number of discoveries, is another performance indicator of practical importance. For specific combinations of $\beta_j$ and $\gamma_j^*$, scatterplots were produced in order to compare random outcomes of the two procedures. The results were largely similar when the training sample was used to assess the performance of $Bonf^\gamma$ and $BH^\beta$.

Another way to assess the performance of the extended Bonferroni procedure is to apply it to a large-scale "spike-in" microarray dataset where the identities of all null and alternative hypotheses are known exactly. However, no high quality datasets of this type are currently available. We explored the possibility of using a recently published "spike-in" control dataset for the Affymetrix Drosophila high-density oligonucleotide arrays [Choe et al. (2005)] for this purpose but found it to be of little value. In addition to the extremely small sample size ($n = 3$ per group), the experimental design behind the study by Choe et al. appears to have many flaws [see Dabney and Storey (2006) for discussion]. Another point that should be mentioned is that spike-in data do not represent a good experimental model of the actual correlation structure of microarray data. These are the reasons why we used real microarray data on a group of patients with childhood leukemia [Yeoh et al. (2002)] to simulate an artificial spike-in dataset, while preserving the actual correlation structure in the same way as defined by the subset pivotality condition [Westfall and Young (1993)]. The description of this data set and its analysis is given in "Supplementary Material 1."

**4. Analysis of simulated data.** The requirements imposed on $Bonf^\gamma$ and $BH^\beta$ were made comparable by equalizing their estimated true error rates (FDR or PFER) as described in Section 3. A training set of 500 pairs of simulated vectors was used for this purpose. The performance of both procedures was assessed on an independently generated control set of 500 pairs of such vectors. Presented in this section are the results obtained by equalizing the estimated FDR, while similar results pertaining to the PFER are given in "Supplementary Material 2." The multiple testing procedure $BH^\beta$ was applied to the control sample with various thresholds $\beta_j$, following which the adjusted thresholds $\gamma_j^*$ were used to run $Bonf^\gamma$ on the same sample. In order to make sure that the equalizing procedure does a good job, we first compared the mean power of the two procedures at adjusted parameter values. As one would expect, both procedures have, on average, the same power when their estimated FDRs are approximately equal (Figures 1A and 1B). This pattern was observed in both independent and correlated data. The same holds true for the median values (Figures 1C and 1D).

Since the mean lengths of the lists of rejected hypotheses are forced to be equal by the equalizing procedure, we should focus on other characteristics of the random outcomes of both procedures. The cut-offs of both procedures are functions of observed $p$-values and can be quite dissimilar across samples even if their mean values are the same. Therefore, our focus is on the variability of those outcomes that can be observed in simulation experiments. The standard deviation of the number of true positives appears to be higher for the BH procedure than for $Bonf^\gamma$ (Figures 2A and 2B). This advantage of $Bonf^\gamma$ is also seen when its stability is assessed in terms of the standard deviation of the total number of rejected hypotheses (Figures 2C and 2D). Scatterplots of the total number of discoveries for specific combinations of adjusted parameters show a high correlation between outcomes of the two MTPs. Shown in Figure 3A is one such scatterplot for independent data. Closed circles represent those pairs of outcomes that occur in at least ten simulated samples. The equalizing value of $\widehat{FDR}$ is 0.0399 with approximately the same standard error of 0.0008 for both procedures. In 425 out of the 500 simulated samples, both procedures resulted in an identical number of rejected hypotheses. The Bonferroni procedure rejected more hypotheses than the BH procedure in 34 samples, while the reverse situation was observed in 41 samples. The sample correlation coefficient between the numbers of rejected hypotheses is equal to 0.991 for the data presented in Figure 3A.

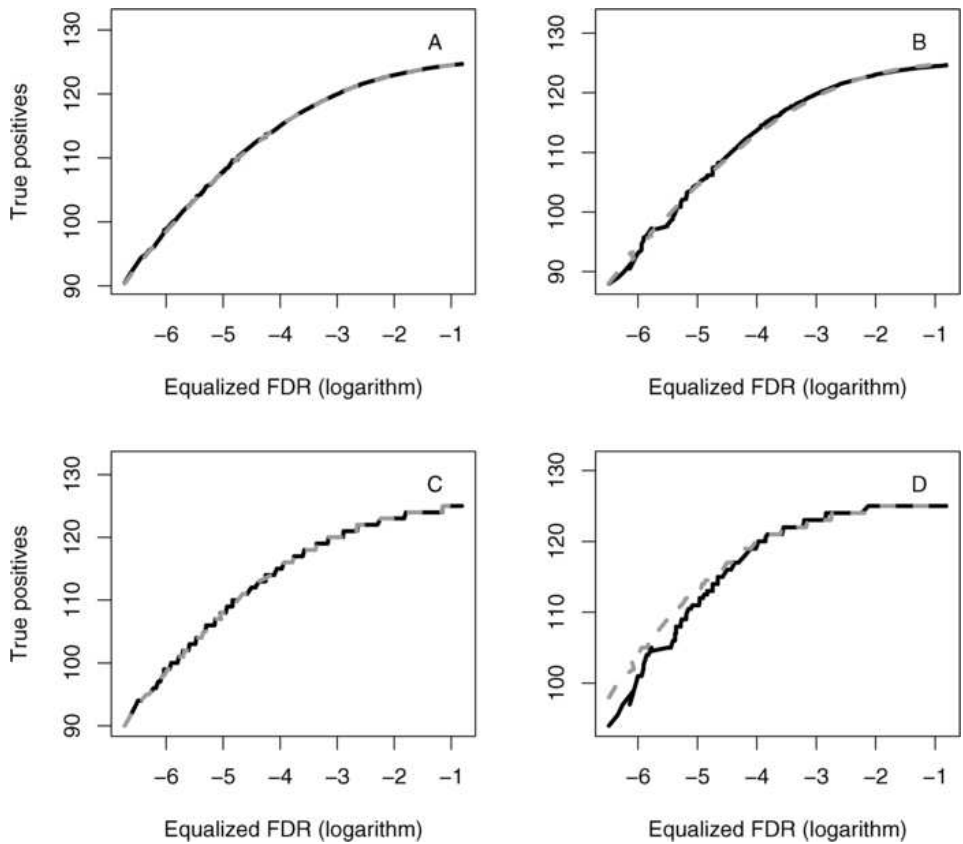

FIG. 1. *The mean (A, B) and median (C, D) of the number of true discoveries produced by $Bonf^\gamma$ and $BH^\beta$ with $\widehat{FDR}$ as the equalizer. A—independent data, B—correlated data, C—independent data, D—correlated data, solid line: $Bonf^\gamma$, dashed line: $BH^\beta$.*

The effect of correlations on the performance of both procedures is quite strong. The standard deviation of the number of true discoveries produced by both procedures is about 5 times larger in the presence of correlation, as moderate as it is in our experiments, than in the independent case (Figures 2A and 2B). The same applies equally to the total number of discoveries (Figures 2C and 2D).

Figure 3B displays the corresponding scatterplot for correlated data. The equalizing value of $\widehat{FDR}$ is 0.0385 with the standard errors of 0.0031 and 0.0033 for $Bonf^\gamma$ and $BH^\beta$, respectively. In 331 out of the 500 simulated samples, both procedures resulted in an identical number of rejected hypotheses. The Bonferroni procedure rejected more hypotheses than the BH procedure in 145 samples, while the reverse situation was observed only in 24 samples. However, as evidenced from Figure 3B, when $BH^\beta$ and $Bonf^\gamma$ differ it is often that $BH^\beta$ gives "much larger" numbers of rejected hypotheses. The correlation coefficient between the numbers of rejected hypotheses is equal to 0.979 in this case.

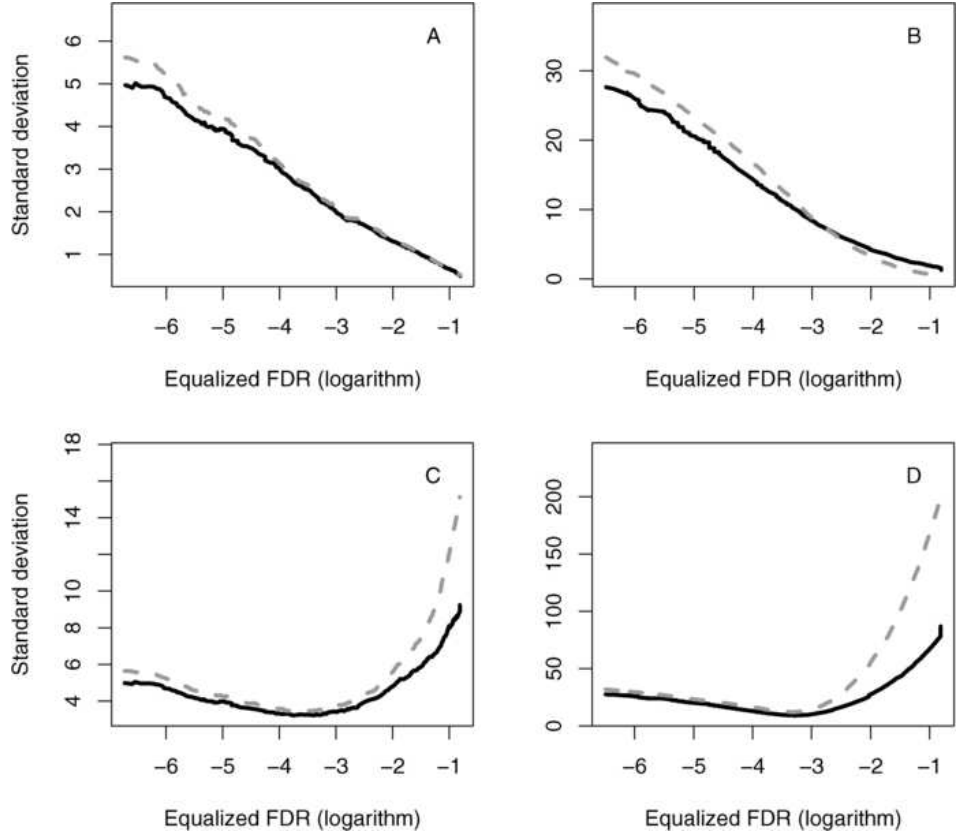

Fig. 2. *The standard deviation of true discoveries (A, B) and total discoveries (C, D) produced by $Bonf^\gamma$ and $BH^\beta$ with $\widehat{FDR}$ as the equalizer. A—independent data, B—correlated data, C—independent data, D—correlated data, solid line: $Bonf^\gamma$, dashed line: $BH^\beta$.*

The results obtained by using $\widehat{PFER}$ as the equalizer are quite similar as shown in "Supplementary Material 2."

**5. Discussion.** Since $Bonf^\gamma$ and $BH^\beta$ are designed to control the conceptually different error rates, it is difficult to compare their power by fixing the "test size" as in the traditional single-test situation. One way around this difficulty is to choose such $\gamma$ and $\beta$ that enforce the same expected level of Type 1 errors for both procedures in accordance with either of the two concepts of error rate. The true error rates (FDR or PFER) for prechosen $\gamma$ and $\beta$ are unknown and can be estimated only if the "true" and "false" null hypotheses are known exactly, which is the case either in simulations or in model experiments such as spike-in microarrays. In the reported study, we used nonparametric estimates of the FDR and PFER as the equalizers to compare higher order characteristics of the power of $Bonf^\gamma$ with those of $BH^\beta$. The same expedient can be employed in power comparisons of any alternative multiple testing procedures.

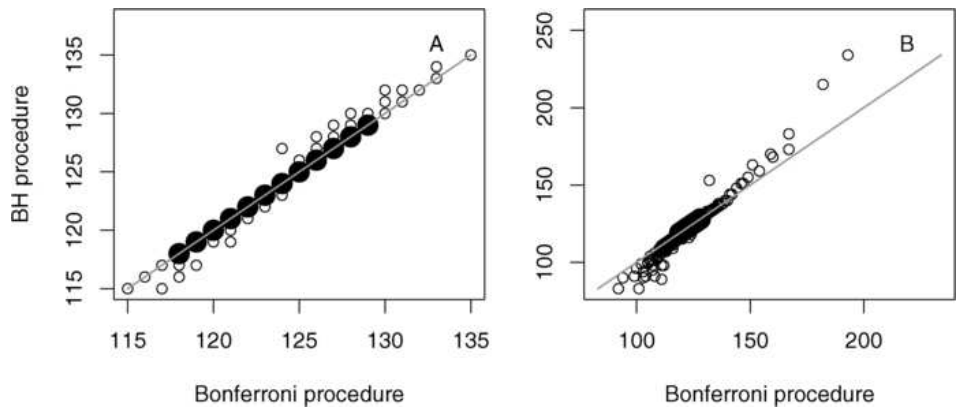

FIG. 3. *The scatterplot of the total number of discoveries for $Bonf^\gamma$ versus $BH^\beta$ with $\widehat{FDR}$ as the equalizer.* A—*independent data,* B—*correlated data, solid line: $Bonf^\gamma$, dashed line: $BH^\beta$. Closed circles represent those outcomes that occur in at least ten simulated samples.*

The results of our study show that the extended Bonferroni procedure $Bonf^\gamma$ can be made as powerful as the $BH^\beta$ procedure by a proper choice of its parameter, a conclusion consonant with that made by Korn et al. (2004). Investigators may prefer one or the other procedure depending on what qualities they perceive as being more important. A distinct advantage of the BH procedure is that its parameter $\beta$ is dimensionless as it represents an upper bound for the expected *proportion* of false positives. Therefore, there is no need to change this parameter when the total number $m$ of hypotheses changes. This makes the BH procedure intuitively appealing. In contrast, the PFER is not scale invariant as it is measured in "items" (hypotheses). Hence, it may be a problem to choose the parameter $\gamma$ for $Bonf^\gamma$ unless an investigator has a clear idea of how many, on average, false positives can be tolerated. In other words, the two procedures serve different purposes. If, for example, the practitioner decides that, on the average, he/she can afford two false positives per experiment, then it is natural to use the Bonferroni procedure with the nominal level of the PFER equaling 2. If, on the other hand, the practitioner wants, on the average, the proportion of false positives among all positives not to exceed 10%, he/she can use the BH procedure with the nominal level of the FDR equaling 0.1.

The main virtue of $Bonf^\gamma$ is its higher stability in terms of the variance of the total number of discoveries, a property that becomes increasingly important with the strength of correlation in the data [Qiu, Klebanov and Yakovlev (2005), Qiu et al. (2006)]. Yet another advantage of $Bonf^\gamma$ is its simplicity. This procedure provides strong control of the PFER at the nominal level $\gamma$ under an arbitrary dependency structure of individual $p$-values. Furthermore, the nominal level is attained under the complete null hypothesis. As for $BH^\beta$, the only proven fact of this nature is that the FDR is controlled to be less than or equal to $\beta$ under the condition of positive regression dependence [Benjamini and Yekutieli (2001)]. In the general case, only a much more conservative version of $BH^\beta$ is available [Benjamini and Yekutieli (2001)].

A very interesting observation is that the standard deviation of the total number of rejections for both procedures has a minimum when considered as a function of the corresponding threshold parameter (Figures 2C and 2D). The minimum is attained when the mean number of rejections becomes close to the total number of true alternative hypotheses. This is not an entirely unexpected phenomenon if one thinks loosely of testing outcomes as a Bernoulli trial. What is important, however, is that the position of this minimum does not change much in the presence of correlations. The curves shown in Figure 2D can be estimated from real microarray data by resampling [Qiu, Klebanov and Yakovlev (2005)] and may be instrumental in estimating a minimal number of differentially expressed genes. This idea and its practical implications invite a special investigation.

A. Gordon
Department of Biostatistics
  and Computational Biology
University of Rochester
601 Elmwood Avenue
Box 630
Rochester, New York 14642
and
Department of Mathematics and Statistics
University of North Carolina at Charlotte
9201 University City Boulevard
Charlotte, North Carolina
USA
E-mail: aygordon@email.uncc.edu

G. Glazko
X. Qiu
A. Yakovlev
Department of Biostatistics
  and Computational Biology
University of Rochester
601 Elmwood Avenue
Box 630
Rochester, New York 14642
USA
E-mail: yakovlev@bst.rochester.edu